\documentstyle[11pt,newpasp,twoside,psfig]{article}
\def\edcomment#1{\iffalse\marginpar{\raggedright\sl#1\/}\else\relax\fi}
\reversemarginpar

\begin{document}
\title{X-ray iron line variability: constraints on the inner accretion disk}
 \author{Christopher S. Reynolds}
\affil{JILA, University of Colorado, Campus Box 440, Boulder, CO~80309}

\begin{abstract}
After reviewing the basic physics of X-ray reflection in AGN, we present
three case studies which illustrate the current state of X-ray
reflection studies.  For the low-luminosity AGN NGC~4258, we find that
the iron line is much narrower than is typically found in higher
luminosity AGN.  We argue that this is evidence for either a truncated
cold accretion disk (possibly due to a transition to an advection
dominate accretion flow at $r\sim 100\,GM/c^2$) or a large ($r\sim
100\,GM/c^2$) X-ray emitting corona surrounding the accretion disk.  We
also present results for the higher luminosity Seyfert nuclei in
NGC~5548 and MCG--6-30-15.  In both of these sources, {\it RXTE} shows
that the iron line equivalent width decreases with increasing
luminosity.  Furthermore, the iron line equivalent width is found to be
{\it anticorrelated} with the relative strength of the reflection
continuum, contrary to all simple reflection models.  It is proposed
that continuum-flux correlated changes in the ionization of the
accretion disk surface can explain this spectral variability.  Finally,
we address the issue of X-ray iron line reverberation in the light of
these complicating factors.
\end{abstract}

\section{Introduction}

AGN are observed to be copious X-ray emitters.  These X-rays are
thought to originate from the innermost regions of an accretion disk
around a central supermassive black hole.  Since the accretion disk
itself is expected to be an optical/UV emitter, the most likely
mechanism producing the X-rays is inverse Compton scattering of these
soft photons in a hot and tenuous corona that sandwiches the accretion
disk.  Thus, in principle, the study of these X-rays should allow the
immediate environment of the accreting black hole as well as the
exotic physics, including strong-field general relativity, that
operates in this environment to be probed.

In the past decade, X-ray astronomy has begun to fulfill that promise.
Guided by observations with the {\it Ginga}, ASCA, RXTE and BeppoSAX
satellites, there is a broad consensus that X-ray irradiation of the
surface layers of the accretion disk in a class of AGN known as Seyfert
1 galaxies gives rise to fluorescent K$\alpha$ emission line of cold
iron via the process of ``X-ray reflection''.  Accompanying the iron
line is a reflected continuum possessing a characteristic shape that is
determined by the competing effects of photoelectric absorption and
Compton scattering.

In this contribution, I will review the basic physical processes leading
to the observed X-ray reflection signatures, and the evidence leading us
to believe that they do indeed originate from the innermost regions of
the accretion disk.  We will then discuss three recent observational
campaigns on NGC~4258 (M106), NGC~5548 and MCG--6-30-15.  The purpose of
presenting results from these campaigns is to describe the
successes, and mysteries, resulting from recent studies of X-ray
reflection signatures.

\section{X-ray reflection spectra and broad iron lines}

The basic physics of X-ray `reflection' can be understood by considering a
hard X-ray (power-law) continuum illuminating a semi-infinite slab of cold
gas.  In this context, `cold' is taken to mean that metal atoms are
essentially neutral, but H and He are mostly ionized.  When a hard X-ray
photon enters the slab, it is subject to a number of possible interactions:
Compton scattering by free or bound electrons, photoelectric absorption
followed by fluorescent line emission, or photoelectric absorption followed
by Auger de-excitation.   A given incident photon is either destroyed by
Auger de-excitation, scattered out of the slab, or reprocessed into a
fluorescent line photon which escapes the slab.   

\begin{figure}[t]
\centerline{
\psfig{figure=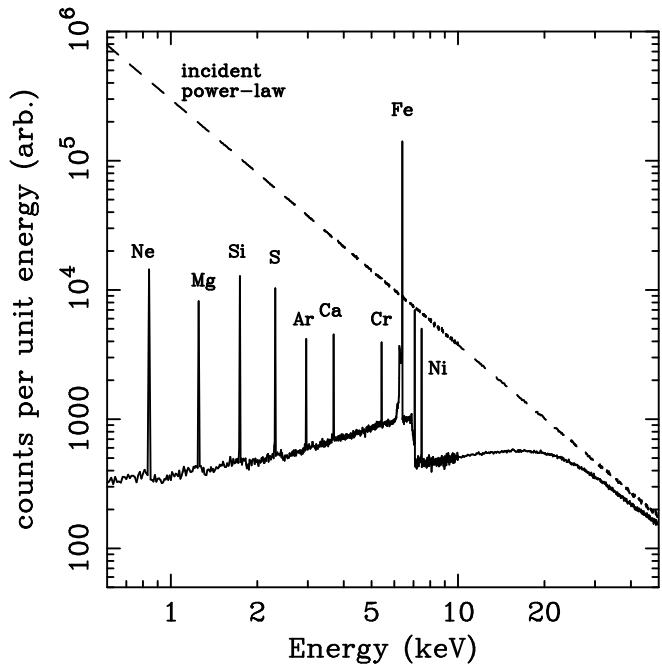,width=1.0\textwidth,angle=270}
}
\caption{X-ray reflection from an illuminated slab.  Dashed line shows
the incident continuum and solid line shows the reflected spectrum
(integrated over all angles).  Monte Carlo simulation from Reynolds
(1996).}
\end{figure}

Figure~1 shows the results of a Monte Carlo calculation which includes all
of the above processes (Reynolds 1996; based on similar calculations by
George \& Fabian 1991).  Due to the energy dependence of photoelectric
absorption, incident soft X-rays are mostly absorbed, whereas hard photons
are rarely absorbed and tend to Compton scatter back out of the slab.  This
gives the reflection spectrum a broad hump-like shape.  In addition, there
is an emission line spectrum resulting primarily from fluorescent K$\alpha$
lines of the most abundant metals.  The iron K$\alpha$ line at 6.4\,keV is
the strongest of these lines.

For most geometries relevant to this discussion, the observer will see this
reflection component superposed on the direct (power-law) primary
continuum.  Under such circumstances, the main observables of the
reflection are a flattening of the spectrum above approximately 10\,keV (as
the reflection hump starts to emerge) and an iron line at 6.4\,keV.  For
solar or cosmic abundances and a plane-parallel slab geometry, the expected
equivalent width of the iron line is 150--200\,eV (George \& Fabian 1991;
Reynolds, Fabian \& Inoue 1995).

\begin{figure}
\centerline{
\psfig{figure=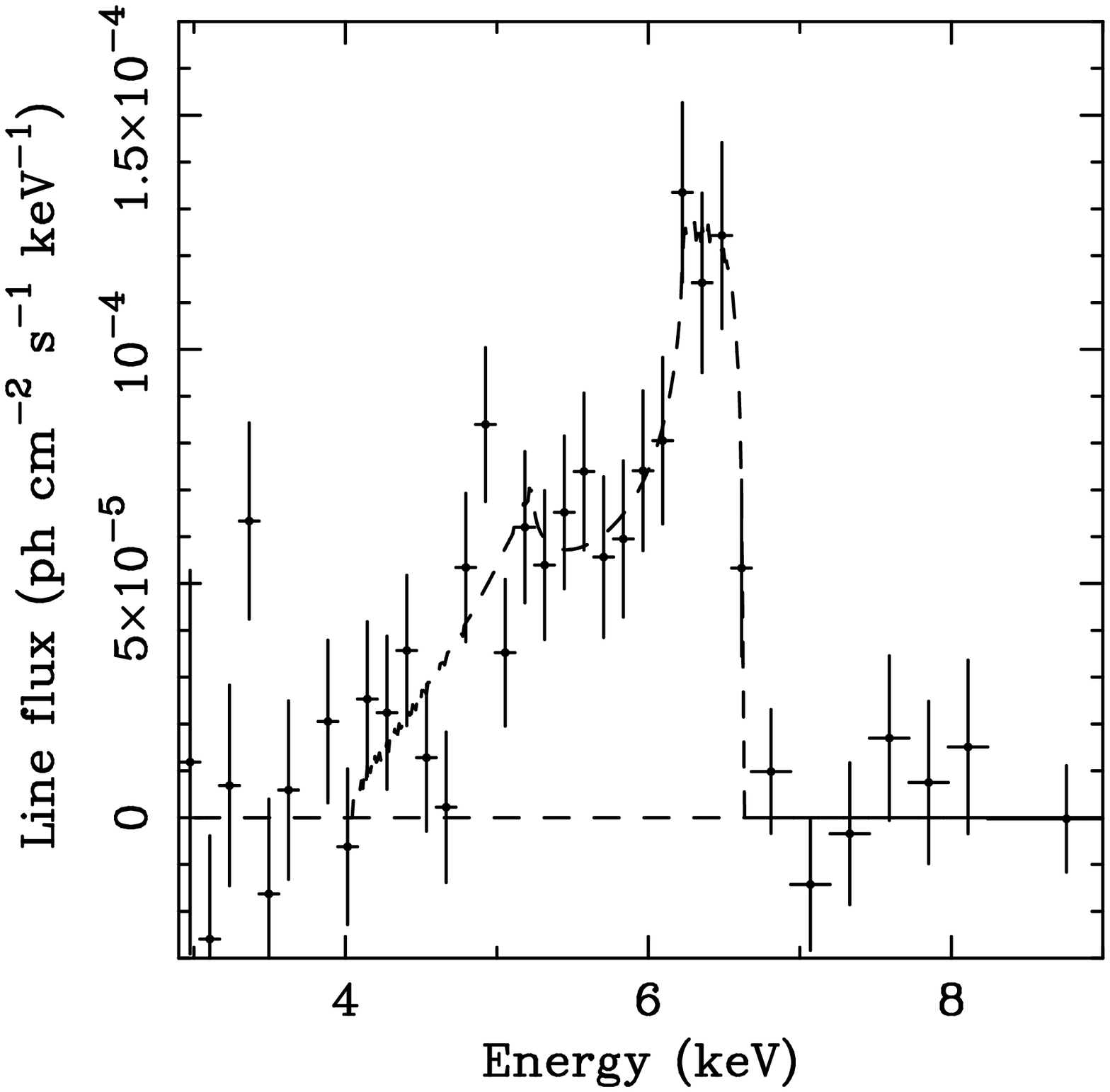,width=0.9\textwidth,angle=270}
}
\caption{Iron line profile from the long 1994 {\it ASCA} observation of
  MCG$-$6-30-15.  From Tanaka et al. (1995).}
\end{figure}

A major advance came with the launch of {\it ASCA} and its
medium-resolution CCD spectrometers.  A long (4.5 day) observation of
the bright Seyfert galaxy MCG$-$6-30-15 in July-1994 allowed the iron
line profile to be determined with some accuracy.  The resulting line
profile is shown in Fig.~2 (published by Tanaka et al. 1995).  The line
was found to be {\it extremely} broad (almost $10^5\,{\rm km}\,{\rm
s}^{-1}$ FWZI) and asymmetric in so far that it possesses an extensive
red-wing.  Such a broad and asymmetric line is expected if the X-ray
reflection is occurring in the inner regions of an accretion disk ---
strong line-of-sight Doppler shifts, transverse Doppler shifts and
gravitational redshifts combine to produce an extensive low-energy wing
and a sharp truncation of the line at high-energies (Fabian et
al. 1989).  Tanaka et al. (1995) showed that the MCG$-$6-30-15 data is
in good agreement with the disk model provided the inclination of the
disk is $\theta\approx 27^\circ$ and line fluorescence occurs down to
$6\,GM/c^2$, the innermost stable orbit around a Schwarzschild
black hole.  {\it This result is the best evidence to date for a
radiatively-efficient accretion disk around a black hole in any object.}

Subsequent studies of large samples of objects by Nandra et al. (1997)
confirmed the presence of these features in many other Seyfert 1
galaxies, and show that there is a tendency for the iron lines to
indicate face-on accretion disks (as expected for Seyfert 1 nuclei from
the unified Seyfert scheme).

\section{Alternative models for broad iron lines}

The claim that iron line studies are probing the region within a few
gravitational radii of the black hole is a bold one, and should be
tested against other models at every opportunity.  Given the quality of
data, the July-1994 MCG$-$6-30-15 line profile has become a test bed for
such comparisons.

Fabian et al. (1995) examined many alternative models including lines
from mildly relativistic outflows, the effect of absorption edges on the
observed spectrum, and broadening of the line via comptonization.
Fabian et al. found that none of these models were viable alternatives
for the MCG$-$6-30-15 line profile.  The idea of producing the broad
line via Comptonization has been revived recently by Misra \& Kembhavi
(1997) and Misra \& Sutaria (1999).  They argue that the spectrum
initially consists of a narrow iron line superposed on a power-law
continuum and that Comptonization in a surrounding cloud with optical
depth $\tau\sim 4$ produces the broad line.  The Comptonizing cloud
must be both cold ($kT<0.5$\,keV in order to predominately
downscatter rather than upscatter the line photons), and fully-ionized
(since no strong iron absorption edges are seen in the continuum
spectrum).  The cloud is kept fully ionized and yet cool by postulating
that the continuum source has a very luminous optical/UV component.

There are strong arguments against such a model.  Since the power-law
continuum emission also passes through any such Comptonizing cloud, one
would observe a break in the continuum spectrum at $E_{\rm br}\sim
m_{\rm e}c^2/\tau^2\sim 30$\,keV.  Such a break is not observed in the
{\it BeppoSAX} (Guainazzi et al. 1999) or {\it RXTE} data (Lee et
al. 1999) for MCG--6-30-15 (see Misra 1999).  Also, both continuum
variability (which is seen on timescales as short as 100\,s) and
ionization arguments limit the size of the Comptonizing cloud in
MCG--6-30-15 to $R<10^{12}$\,cm.  The essence of this ionization argument
is that the ionization parameter at the outer edge of the cloud (which,
for a fixed cloud optical depth, scales with cloud size as $\xi\propto
1/R$) must be sufficiently high that all abundant metals, including
iron, are fully ionized (Fabian et al. 1995; Reynolds \& Wilms 2000).
In the case of MCG--6-30-15, these constraints on the cloud size turn
out to so tight that the postulated optical/UV component required to
Compton cool the cloud would violate the black body limit (Reynolds \&
Wilms 2000). Comptonization moreover provides a poor fit (Ruszkowski
\& Fabian 2000).  Hence, we consider the Comptonization model for the broad
iron line not to be viable.

In another alternative model, Skibo (1997) has proposed that energetic
protons transform iron in the surface of the disk into chromium and
lower $Z$ metals via spallation which then enhances their fluorescent
emission.  With limited spectral resolution, such a line
blend might appear as a broad skewed iron line.  This model suffers both
theoretical and observational difficulties.  On the theoretical side,
high-energy protons have to be produced and slam into the inner
accretion disk with a very high efficiency (Skibo assumes $\eta=0.1$ for
this process alone).  On the observational side, it should be noted that
the broad line in MCG--6-30-15 (Tanaka et al 1995) is well resolved by
the {\it ASCA} SIS (the instrumental resolution is about 150~eV at these
energies) and it would be obvious if it were due to several separate and
well-spaced lines spread over 2~keV. 

\section{Case study I --- the low luminosity AGN NGC~4258}

Low luminosity AGN (LLAGN) are almost certainly systems which are
accreting at very low Eddington fractions.  This places them in a regime
of accretion which has been the subject of active theoretical work for
many years.  It was realized by several authors that when the accretion
rate is low (relative to the Eddington rate), an accretion disk may
switch into a hot, radiatively-inefficient mode (Ichimaru 1977; Rees
1982; Narayan \& Yi 1994; Narayan \& Yi 1995).  In essence, the plasma
becomes so tenuous that the timescale for energy transfer from the
protons to the electrons (via Coulomb interactions) becomes longer than
the inflow timescale.  The energy remains as thermal energy in the
protons (which are very poor radiators) and gets advected through the
event horizon of the black hole.  These are the so-called Advection
Dominated Accretion Flows (ADAFs\footnote{In recent years there has been
a large amount of work on advective accretion flows, with a
corresponding expansion in terminology and acronims (e.g. ADIOS, BDAF,
and CDAF). See Prof.~Narayan's contribution in these proceedings for a
fully discussion of the various types of advective accretion flow.
Here, we use the term ADAF to encompass all such accretion flows which
are radiatively inefficient.}).  ADAFs are to be contrasted with
`standard' radiatively-efficient accretion disks in which the disk
remains cool and geometrically thin all of the way down to the black
hole (Shakura \& Sunyaev 1973; Novikov
\& Thorne 1974).

LLAGN provide an ideal laboratory in which to study ADAFs.  If the
central regions of LLAGN accretion disks are operating in an ADAF mode,
we would not expect to observe broad iron lines since the ADAF is far
too hot to contain combined iron ions.  On the other hand, if we see
broad iron lines from a LLAGN, this would be strong evidence for radiatively
efficient disks even at these low accretion rates.  However, the fact
that LLAGN are faint has hampered such detailed X-ray studies of these
sources.

\begin{figure}
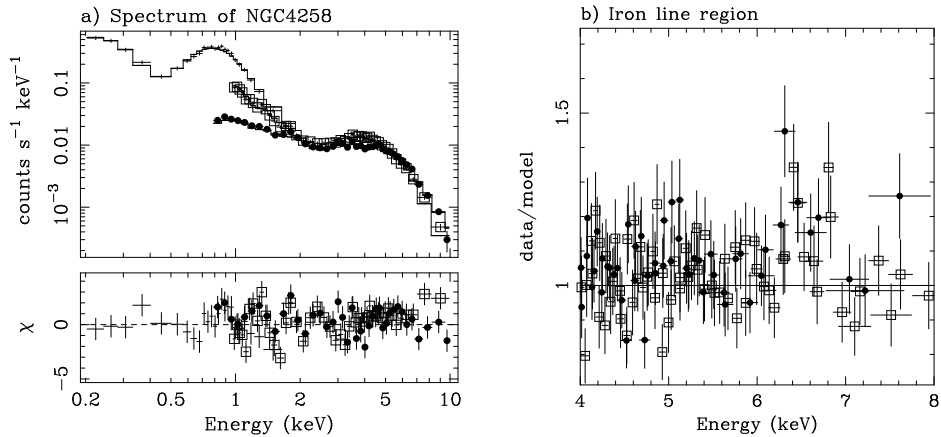

\hbox{
\psfig{figure=ngc4258_spec.ps,width=0.45\textwidth,angle=270}
\hspace{0.5cm}
\psfig{figure=ngc4258_ironline.ps,width=0.42\textwidth,angle=270}
}
\caption{Panel (a) shows the joint {\it ROSAT}/{\it ASCA} spectrum for
NGC~4258 fit with model-F from Table~1.  The {\it ROSAT} data are the
plain crosses.  For clarity, only data from the SIS0 (open squares) and
GIS2 (filled circles) instruments on board of {\it ASCA} are shown.
Panel (b) shows iron line region of the spectrum with less severe
binning, also referenced to model-F from Table~1.  The presence of a
fluorescent iron line is clear.  These are both folded spectra (in the
sense that they include the instrumental response).}
\end{figure}

In an attempt to study one LLAGN in some detail, we performed a
200\,ksec joint {\it ASCA}/{\it RXTE} observation of the nearby LLAGN
NGC~4258.  A full discussion of this observation is presented in
Reynolds, Nowak \& Maloney (2000).  Here, we summarize the iron line
properties of this object.  

As can be seen in Fig.~3, we clearly detect a iron line with an
equivalent width of $W_{K\alpha}=107^{+42}_{-37}$\,eV and a velocity
width of $<22000\,{\rm km}\,{\rm s}^{-1}$ (FWHM) which is much narrower
than those seen in higher luminosity Seyfert galaxies.  If it is assumed
that the observed iron line originates from the accretion disk, then we
can use relativistic ``disk-line'' models to place constraints on the
location of the line emission.  Using such models it is found that the
narrowness of the line implies that the bulk of the line emission
originates from a distance $r>100\,GM/c^2$ from the black hole.  Such a
large line emitting radius is consistent with the ADAF model for
NGC~4258 (Lasota et al. 1996) in which the thin fluorescing accretion
disk goes through an transition into an ADAF at about $r\sim
100\,GM/c^2$.  However, the data are also consistent with a radiatively
efficient fluorescing disk extending down to the radius of marginal
stability and a large ($r\sim 100\,GM/c^2$) X-ray corona.  In this
latter case, there would be faint broad wings to the line profile which
could be found with a high signal-to-noise {\it XMM--Newton} spectrum.
One way or another, the iron line properties of this LLAGN are different
to those of many higher-luminosity Seyfert galaxies which display a
broad iron line indicating radiative disks and small ($\sim 10\,GM/c^2$)
X-ray emitting regions.

\begin{figure}[t]
\centerline{
\psfig{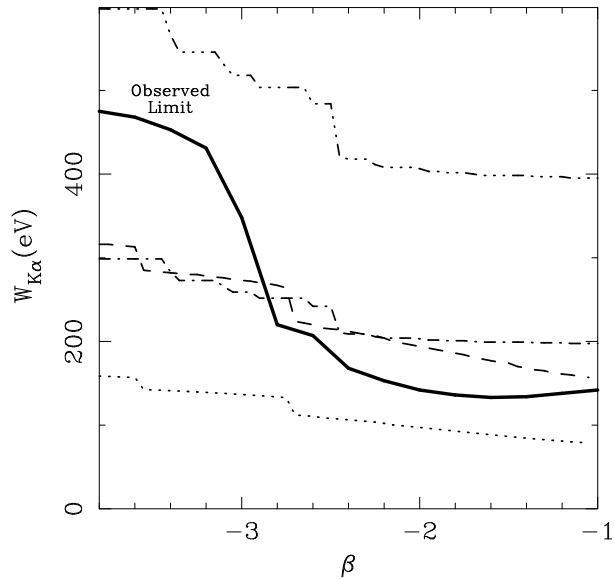}
}
\caption{Constraints on the presence of a ``Seyfert-like'' iron line in the 
case where the observed narrow line is modeled by a separate narrow
Gaussian component.  The broad line is modeled as originating in an
accretion disk around a Schwarzschild black hole, with a disk
inclination of $i=85^\circ$ and an inner line emitting radius of
$r_{\rm br}=6GM/c^2$.  The solid line shows the upper limit on the
equivalent widths a function of the emissivity index $\beta$.  The
dotted and dashed lines show the theoretical expectation, taking into
account limb-darkening and light bending effects, assuming that the iron
line has an equivalent width of $200$\,eV and $400$\,eV, respectively,
when the disk is viewed face on.  The dot-dashed and dot-dot-dot-dashed
lines show the theoretical expectation when limb-darkening is absent
(see text) assuming that the iron line has an equivalent width of
$200$\,eV and $400$\,eV, respectively, when the disk is viewed face on.}
\end{figure}

We cannot rule out the possibility that most (or all) of the observed
iron line originates from matter that is not associated with the
accretion disk.  However, simple arguments lead us to disfavor such a
scenario.  Consider iron line emission in a geometrically-thick torus
surrounding the accretion disk of NGC~4258.  An upper limit to the
column density of this structure along our line of sight to the AGN is
given by the observed column density of $N_{\rm H}\approx 10^{23}\,{\rm
cm}^{-2}$.  If we suppose that this torus is in the same plane as the
accretion disk (so that we are also viewing it edge-on), it is plausible
to assume that we are looking through the optically-thickest part of the
torus.  We can then show that the maximum equivalent width produced by
fluorescence of this material is $W_{\rm Fe, max}\approx 65$\,eV
(Reynolds, Nowak \& Maloney 2000).  However, there are several reasons
why this upper limit would almost certainly not be achieved.  Firstly,
modeling of the accretion disk warp strongly suggests that our line of
sight intercepts the disk and that the bulk of the column density which
obscures the AGN originates in the disk (Herrnstein,
priv. communication).  Therefore, we might expect significantly smaller
column densities along lines of sight that have smaller inclinations
angles relative to the accretion disk.  Secondly, the accretion disk may
well occult half of this fluorescing cloud, thereby reducing this
prediction further.  Thus, the true iron line from surrounding non-disk
material may well be reduced from our naive prediction by a factor of
several.

Despite these arguments, let us now suppose that the observed iron line
is {\it not} associated with the accretion disk.  Our data then permit
the presence of a ``Seyfert-like'' broad iron line which is buried in
the noisy continuum data.  If we suppose NGC~4258 to possess a normal
``Seyfert-like'' broad iron line (with an intrinsic equivalent width in
the range 200--400\,eV, modified by the light bending and limb-darkening
effects that are important in edge-on systems such as NGC~4258), we can
set some interesting constraints (see Fig.~4).  In particular, we can
rule out models in which the limb-darkening is very weak (as would be
the case if, for example, the $\tau_{\rm e}=1$ surface of the accretion
disk is very filamentary rather than planar).  Of course, higher
signal-to-noise data from {\it XMM--Newton} are required if one is to
constrain the nature of the disk in this object any further.

\section{Case study II --- an EUV/X-ray campaign on NGC~5548}

In the rest of this contribution, we turn to the properties of higher
luminosity systems.  Firstly, we shall discuss the classical Seyfert
galaxy NGC~5548.

We\footnote{The team of investigators consisted of Omer Blaes (UCSB),
James Chiang (Colorado), Greg Madjeski (NASA-GSFC), Pawel Magdziarz
(deceased), Hermann Marshall (MIT-CSR), Mike Nowak (Colorado) and
myself.} observed NGC~5548 with {\it EUVE}, {\it ASCA}, and {\it RXTE}
in June/July 1998.  The main part of the campaign consisted of four
observations (with all of these satellites simultaneously) separated by
approximately one week.  One of these observations (18-June-1998) was
substantially longer than the others ($\sim 110$\,ksec of good data, as
compared with an average of $\sim 30$\,ksec for the other observations).
This long observation provided us with intensive monitoring over a
period of 3 days from the EUV to hard X-rays.  The various aspects of
this campaign are discussed in detail in Chiang et al. (2000).

\begin{figure}[t]
\centerline{
\psfig{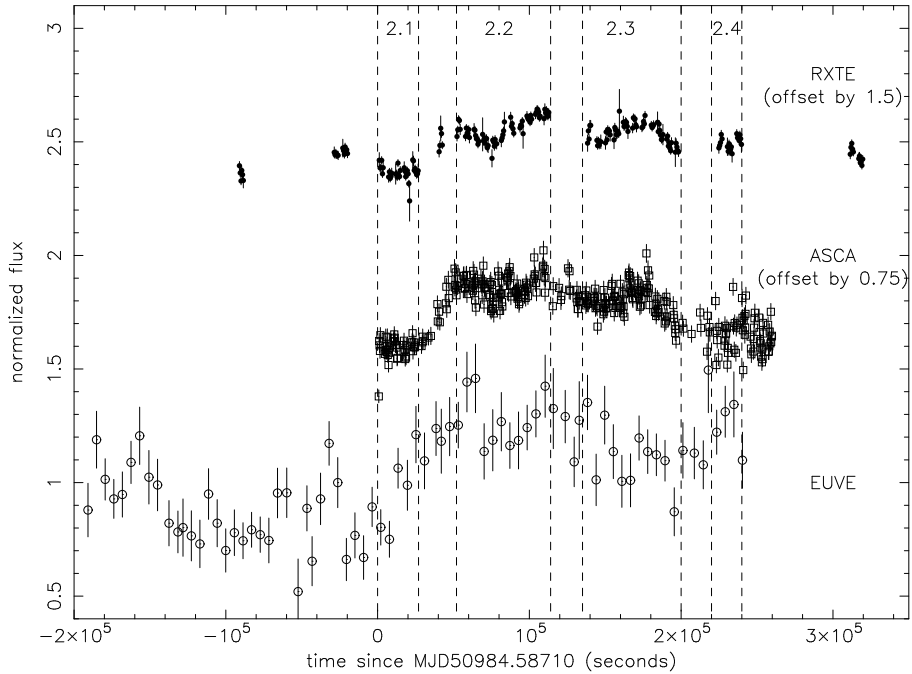}
}
\caption{{\it EUVE}, {\it ASCA}-SIS, and {\it RXTE}-PCA light curves for
the long observation of NGC~5548.  All light curves have been normalized
to a unity mean level.  For clarity, the {\it ASCA} and {\it RXTE}
light curves have then been offset by 0.75 and 1.5, respectively.}
\end{figure}

Although not strictly related to X-ray reflection studies, it is worth
mentioning the continuum variability properties found during this
campaign.  Figure~5 shows the {\it EUVE}, {\it ASCA} and {\it RXTE}
light curves for our long observation.  The most distinct feature is a
step at $\sim 2\times 10^4$\,s.  By computing the z-transformed discrete
correlation function (ZDCF; Alexander 1997) between these various data,
we find clear evidence that the EUV emission leads the {\it RXTE}-band
emission by $\sim 30$\,ksec, and the {\it ASCA}-band emission by $\sim
15$\,ksec.  If confirmed, this results has at least two important
implications.  Firstly, the EUV emission cannot simply be reprocessed
hard X-ray emission since these scenario would predict the EUV to lag
the hard X-rays.  Secondly, if we make the assumption that the observed
lags are due to the Compton upscattering of seed photons, we can
constrain the size of the Comptonizing region to be about $10\,GM/c^2$
for a $10^8M_\odot$ black hole.

\begin{figure}[t]
\centerline{
\psfig{figure=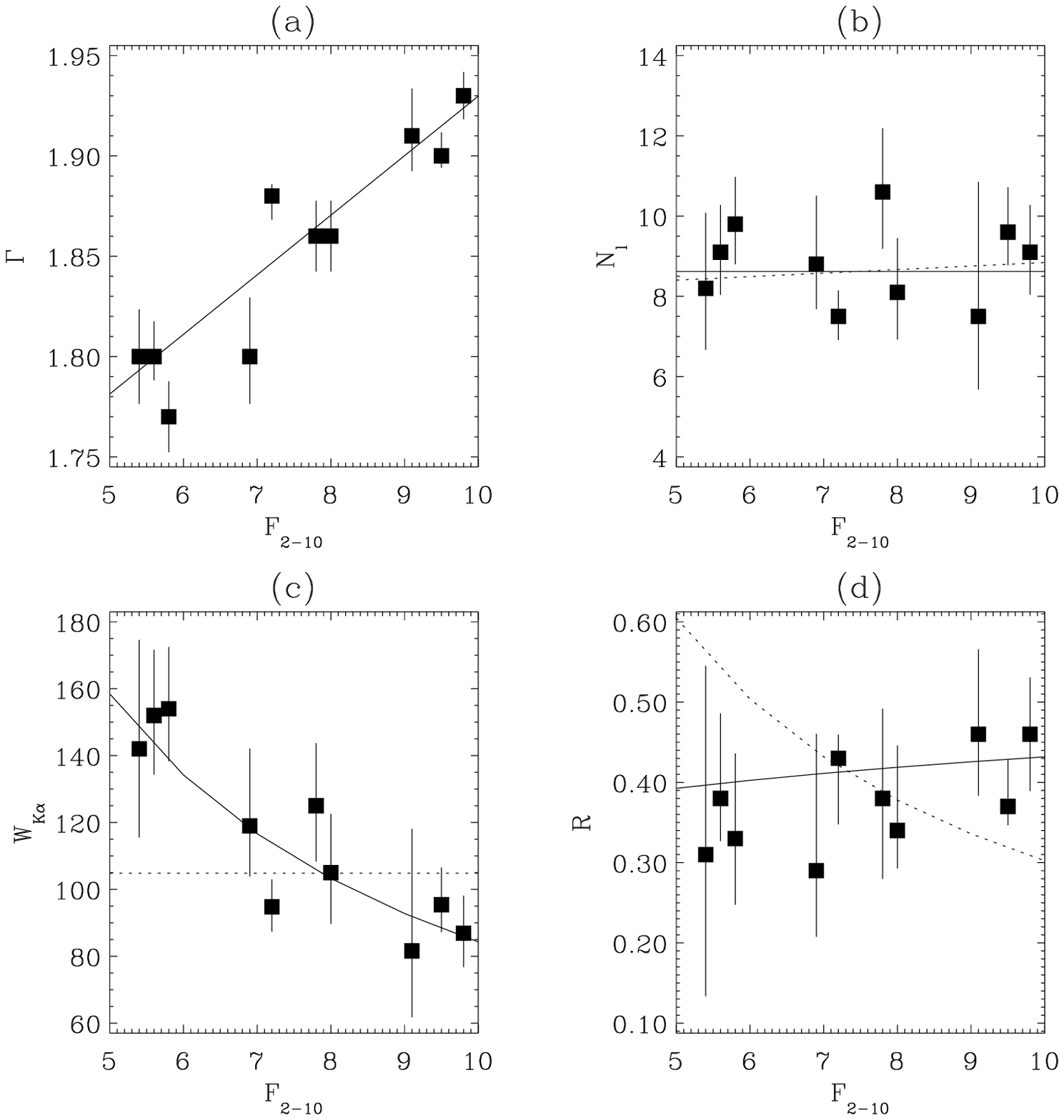,width=0.7\textwidth}
}
\caption{{\it RXTE} results on NGC~5548.  Panel (a) shows the 
well-known trend for the X-ray photon index to increase slightly as the
continuum flux increases (possibly due to Compton cooling of the
corona).  Panels (b) and (c) show that the iron line possesses
approximately constant flux (the dotted line in panel-c shows the best
fitting constant).  Panel (d) shows that the relative strength of the
reflection continuum is approximately constant or slightly increasing.
The dotted line shows the predicted strength of the reflection continuum
assuming that it tracks the iron line equivalent width.  Figure from
Chiang et al. (2000).}
\end{figure}

Returning to the issue of X-ray reflection, this campaign produced a
major surprise.  Before discussing the data, it is worth recapping our
theoretical expectations from simple X-ray reflection models.  For a
fixed disk/corona geometry and ionization state, we expect the iron line
equivalent width to remain constant provided one is probing timescales
longer than the reverberation timescale of the system (which is always
the case for spectral studies with {\it ASCA} and {\it RXTE}).  If the
iron line equivalent width does change due to, for example, a change in
the geometry of the system, then the relative strength of the reflection
continuum would be expected to vary in step. In other words, we expect a
strict proportionality between the iron line equivalent width and
relative strength of the reflection continuum.

\begin{figure}[t]
\centerline{
\psfig{figure=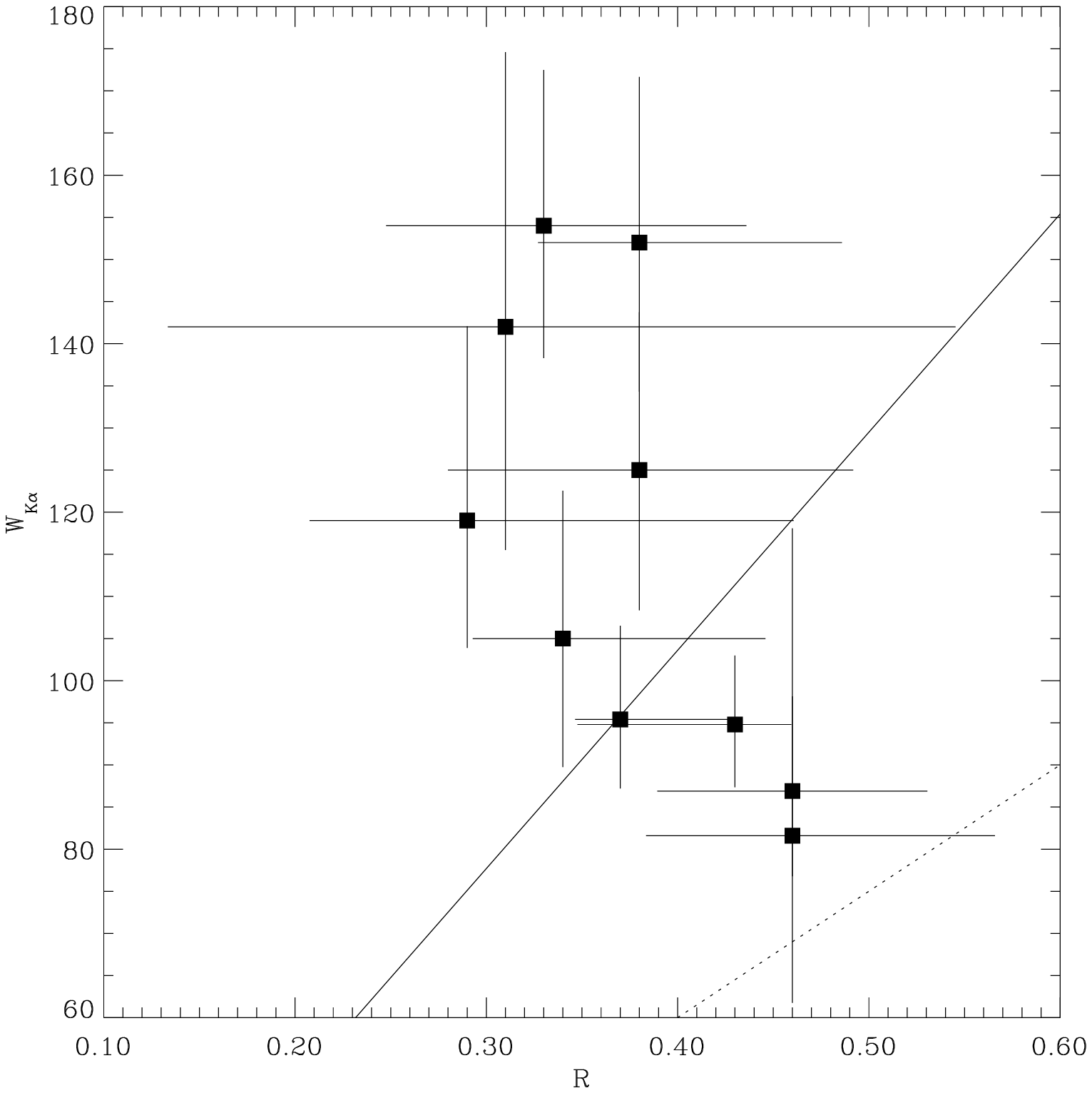,width=0.7\textwidth}
}
\caption{Iron line equivalent width $W_{K\alpha}$ vs. the relative
strength of the reflection continuum ${\cal R}$ for NGC~5548.  The
dotted line is the expected linear relationship assuming solar
abundances for the cold reflector (George \& Fabian 1991).  The solid
line is the best fit proportionality relationship which is still in
conflict with the data at the 89\% level.  Figure from Chiang et
al. (2000).}
\end{figure}

Figure~6 shows the various spectral parameters, measured from the {\it
RXTE} data, for NGC~5548 as a function of the 2--10\,keV continuum flux
of the source.  We found that the equivalent width of the iron line
declined as the continuum source flux increased --- in fact, the iron
line {\it flux} was consistent with being constant.  Simultaneous
spectroscopy with {\it ASCA} showed the iron line to be broadened.
Hence, the line does appear to be originating from the central regions
($r<50\,GM/c^2$) of the accretion disk and, for any reasonable size
black hole mass, the constancy of the line flux cannot be due to light
travel time effects.  Moreover, the relative reflection fraction ($R$)
fails to track the changes in the iron line equivalent width.  Indeed,
these quantities appear to be weakly {\it anti-correlated} (see Fig.~7),
contrary to the simple reflection models.

\section{Case study III --- the X-ray campaign on MGC--6-30-15}

\begin{figure}[t]
\centerline{
\psfig{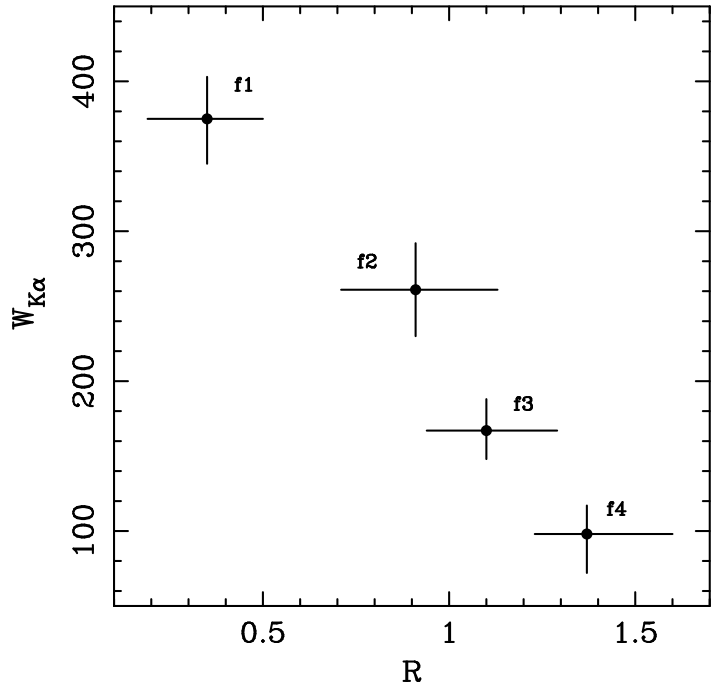}
}
\caption{Iron line equivalent width $W_{K\alpha}$ vs. the relative
strength of the reflection continuum ${\cal R}$ for MCG--6-30-15.   Flux
states f1--f4 are in order of increasing continuum flux.  Figure from
Lee et al. (2000).}
\end{figure}

A very long, simultaneous {\it ASCA}/{\it RXTE} observation of the
Seyfert 1 galaxy MCG--6-30-15 by an (almost) independent team of
researchers\footnote{The team of investigators consisted of Niel Brandt
(PSU), Andy Fabian (Cambridge), Kazushi Iwasawa (Cambridge), Julia Lee
(MIT) and myself.} revealed very similar spectral results to those
obtained for NGC~5548.  Figure~8 shows the clear anti-correlation
between iron line equivalent width and the relative strength of the
reflection continuum found by Lee et al. (2000).  To produce this
figure, almost nine days of {\it RXTE} data have been binned into four
flux-sorted spectra.  Unlike with NGC~5548, these data strongly require
a change in the relative strength of the reflection continuum as well as
the iron line equivalent width.  The details of this observation are
discussed by Lee et al. (1999, 2000).

\begin{figure}[t]
\hbox{
\psfig{figure=mcg6_rec.ps,width=0.45\textwidth,angle=270}
\psfig{figure=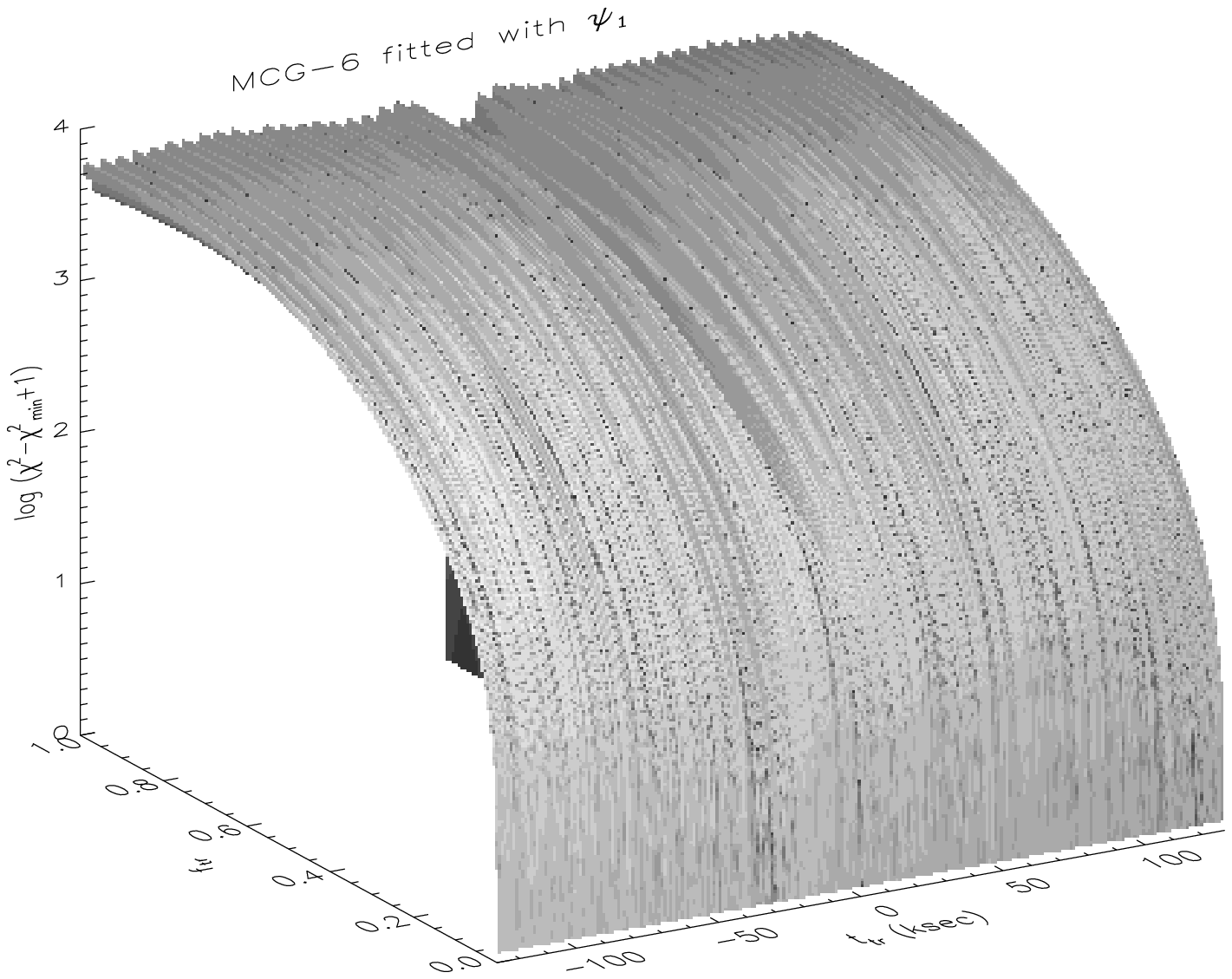,width=0.55\textwidth}
}
\caption{Panel (a) shows 2 orbits of the 2-4\,keV light curve (open
squares and thin error bars) together with the reconstructed light curve
(filled circles and heavy error bars).  Panel (b) shows the $\chi^2$
surfaces and confidence contours resulting from applying trial transfer
function $\psi_1=(1-f_{\rm tr})\delta(t)+f_{\rm tr}\delta(t-t_{\rm tr})$
to the reconstructed continuum light curves and comparing with the line
band light curve.  The surface is plotted using
$log_{10}(\chi^2-\chi^2_{\rm min}+1)$ as the ordinate in order to
display the topography of the region near the minimum.  The $\chi^2$
minimum is along the line $f_{\rm tr}=0$ indicating that no time lag has
been detected.  Figure from Reynolds (2000).}
\end{figure}

This particular {\it RXTE} campaign yielded data of an almost
unprecedented nature --- the observation lasted over 100 consecutive
orbits of {\it RXTE}, giving us dense coverage ($>50\%$) for almost 9
days.  Reynolds (2000) took advantage of this unique dataset and
performed a detailed search for delays between a continuum band
(2--4\,keV) and the iron line band (5--7\,keV) in an attempt to search
for iron line reverberation effects.  In essence, the method of Press,
Rybicki \& Hewitt (1992) was used to interpolate across the gaps in the
continuum light curve (caused primarily by Earth occultation) and
estimate the errors in the interpolation.  Figure~9a shows a small
portion of the light curve, together with the Press et al. (1992)
reconstruction.  The continuous (interpolated) continuum light curve can
then be folded through a trial transfer function and compared with the
line-band light curve in a $\chi^2$ sense (e.g., Fig.~9b shows the
$\chi^2$ surface that results from such a procedure).  By varying the
parameters of the trial transfer function, we can place constraints on
any time delays between bands.  No time delays were found on timescales
down to 500--1000\,s.  Instead, it was found that the line flux remained
constant, thereby extending the result of Lee et al. (2000) down to
these timescales.   

\section{Evidence for flux-correlated changes in the disk's ionization}

To summarize the mysteries resulting from the NGC~5548 and MCG--6-30-15
campaigns, it is found that:
\begin{enumerate}
\item The iron line {\it flux}, rather than the equivalent width,
appears to remain approximately constant as the continuum X-ray source
undergoes rapid variability.  In NGC~5548, this result is derived by
direct X-ray spectral fitting and so applies on timescales of $\sim
50$\,ksec (which is approximately timescale that can be probed via such
methods).  In MCG--6-30-15, the spectral fitting of Lee et
al. (2000) combined with the analysis of Reynolds (2000) leads us to
believe that such behavior occurs on all timescale from 500\,ksec
down to 500\,s.  This cannot be due to light travel delays.
\item The iron line equivalent width and relative strength of the
reflection continuum are {\it anti-correlated}, in contradiction to
simple X-ray reflection models.
\end{enumerate}
Changing the ionization state of the surface of the accretion disk is
one of the few ways of breaking the expected proportionality between the
iron line equivalent and the (inferred) relative strength of the
reflection continuum.   

\begin{figure}
\centerline{
\psfig{figure=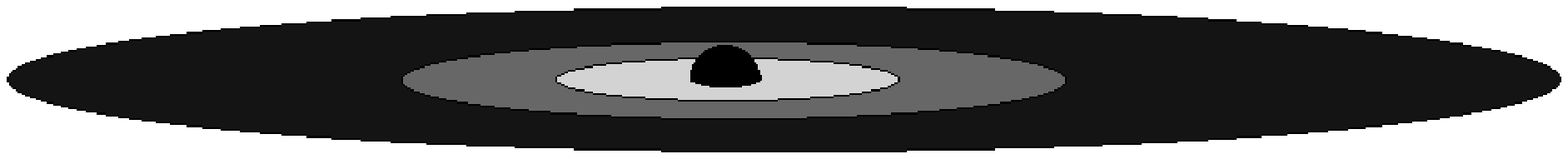,width=0.9\textwidth}
}
\caption{The toy model of an ionized accretion disk that may 
produce the observed spectral variability.  The inner (light) region is
strongly irradiated, highly ionized and does not produce any observable
reflection features.  The outer (dark) region is weakly irradiated,
weakly ionized, and produces `neutral-like' reflection signatures.  The
partially ionized middle zone may be able to produce a significant
reflection continuum with very little iron line (see text).  As the
primary X-ray source varies in intensity, the boundaries between these
zones (presumably defined by some given ionization parameters) will
move, thereby changing the relative strengths of the iron line and
reflection continuum.  See Reynolds (2000) for a more detailed
discussion.}
\end{figure}

Reynolds (2000) proposed a toy model to explain these {\it RXTE} results
(Fig.~10).  In this model, there is a radial ionization gradient on the
disk surface due to the radial dependence of both the ionizing flux and
the density of the surface layers.  The surface of the inner disk, which
is strongly irradiated by the primary X-ray flux, is highly ionized and
acts as a pure Compton mirror.  It thereby fails to imprint distinct
(atomic) features in the observable spectrum.  By contrast, the outer
disk surface (starting at $\sim 20-50\,GM/c^2$) possesses a relatively
low ionization parameter and produces `neutral-like' X-ray reflection
signatures such as those calculated by George \& Fabian (1991).  The
interesting and uncertain processes occur in the partially ionized
intermediate zone of the accretion disk.  In this region, there are two
possible physical mechanisms by which the iron line can be strongly
suppressed relative to the reflected continuum.  These mechanisms are
described in the next two paragraphs.  Reynolds (2000) noted that, if
the boundaries between these zones are defined by some given set of
ionization parameters, the zones will move as the primary continuum flux
varies.  If the primary continuum flux increases, the fluorescing outer
zone will move outwards and so subtends a smaller solid angle as seen by
the primary X-ray source.  The observed iron line equivalent width will
then be seen to decrease.  However, the measured relative strength of
the reflection continuum may remain roughly constant if the intermediate
zone still produces bound-free absorption features in the reflection
continuum.

As mentioned above, there are at least two mechanisms which may
operate in the intermediately ionized zone of the accretion disk to
suppress iron line emission relative to the reflection continuum.
Nayakshin, Kazanas \& Kallman (2000) showed that the well-known
thermal instability of photoionized plasma could lead to the formation
of a highly ionized skin on the surface of the accretion disk.
Compton scattering in this skin would smear out sharp spectral
features such as the iron line, possibly to the level where line
photons from this region would no longer be identified with the iron
line.  However, very broad spectral features such as the iron edge and
broad hump that observationally define the reflection continuum, would
be much less affected by Compton scattering in the skin.  Thus, from
the observational point of view, the presence of such a skin might
reduce the iron line strength while leaving the strength of the
reflection continuum unaffected.    

The second line destruction mechanism is that of resonant trapping
followed by Auger destruction (Ross \& Fabian 1993).  A fluorescent iron
line photon is typically created at an optical depth of $\tau_{\rm
e}\sim 1$.  However, when the bulk of the ions are at least as ionized
as Fe\,{\sc xvii}, the presence of a vacancy in the L-shell of the iron
ions leads to resonant scattering of the line photons with a rather
small path length.  If, furthermore, the bulk of the ions possess at
least two L-shell electrons (i.e. Fe\,{\sc xxiii} or less), then the
Auger effect acts as a photon destruction mechanism.  The combination of
resonant `trapping' and Auger `destruction' might reduce the observed
iron line to a very small strength.

Clearly, more work is needed on the physics and modeling of ionization
accretion disks in order to assess whether such a toy model can explain
the mysterious X-ray variability that we see in current {\it RXTE} data.

\section{Down to the reverberation timescale}

The rapid X-ray variability of many Seyfert galaxies leads us to believe
that the primary X-rays are emitted during dramatic flare-like events in
the accretion disk corona.  When a new flare becomes active, the hard
X-rays from the flare will propagate down to the cold disk and excite
iron fluorescence.  Due to the finite speed of light, the illumination
from the flare sweeps across the disk, and the reflected X-rays act as
an `echo' of this flare.  Such flaring will cause temporal changes in
the iron line profile and strength due to the changing illumination
pattern of the disk and, more interestingly, time delays between the
observed flare and the its fluorescent echo.  This latter effect is
directly analogous to the optical broad line reverberation that is a
major theme of this meeting.  The observational study of iron line
reverberation is beyond the capabilities of current instruments (even,
probably, {\it XMM--Newton}) but should be within reach of future
high-throughput X-ray observatories such as {\it Constellation-X} and
{\it XEUS}.  

In recent work, it has been shown that interesting diagnostics of the
black hole spin and X-ray source geometry are encoded within the iron
line reverberation pattern (Reynolds et al. 1999; Young
\& Reynolds 2000; Ruzskowski 2000).  However, there are many
complications and issues that will affect our ability to extract these
reverberation signatures from real data.  Firstly, there are likely to
be several X-ray flares active at any one time which will give
overlapping reverberation signatures.  Moreover, due to the fact that
the X-ray emitting corona is probably extended on the same spatial
scales as the inner disk, each X-ray emitting flare will have its own
distinct transfer function. This breaks the linearity of the transfer
problem.  Young \& Reynolds (2000) showed that if there are only a small
number of powerful flares, the individual transfer functions can be
identified and separated in {\it Constellation-X data}.  However, we
must examine more powerful statistical methods for extracting useful
information from this system.

Possible reverberation signatures will be affected by short timescale
ionization changes.  The models of Reynolds et al. (1999) and Young \&
Reynolds (2000) assume that the disk has a fixed, fairly cold,
ionization structure.  However, as discussed extensively above, {\it
RXTE} argues for ionization changes in the disk surface that are
correlated with the continuum flux.  On a local scale, this effect may
be driven by ionization fronts that sweep across the disk when an X-ray
flare erupts.  In future work, these effects must be assessed so that
instruments and observing strategies can be planned to study
reverberation effects.  It must be noted that, even though {\it RXTE}
data argue for almost constant iron line flux due to the ionization
changes, we would still expect to see variability of the line strength
and/or profile on either the recombination timescale in the disk surface
or the light-crossing timescale from the flare to the disk, whichever is
longest.

Given all of these complications and difficulties, why bother to pursue
the idea of iron line reverberation?  The answer to this question is
straightforward --- until we have the technology to directly image the
immediate environment of a black hole (and this is at least 20 years
away), X-ray iron line reverberation is simply our best hope of mapping
mapping out the detailed astrophysical environment of supermassive black
holes, including the effects of the black hole's spin.

\section{Conclusions}

In this contribution, we present three case studies in order to describe
the current state of X-ray reflection studies in AGN.   For the
low-luminosity AGN NGC~4258, we find that the iron line is much narrower
than is typically found in higher luminosity AGN.   We argue that this
is evidence for either a truncated cold accretion disk (possibly due to
a transition to an ADAF at $r\sim 100\,GM/c^2$) or a large ($r\sim
100\,GM/c^2$) X-ray emitting corona surrounding the accretion disk.   We
also present results for the higher luminosity Seyfert nuclei in
NGC~5548 and MCG--6-30-15.  In both of these sources, {\it RXTE} shows
that the iron line equivalent width decreases with increasing
luminosity.  Furthermore, the iron line equivalent width is found to be
{\it anticorrelated} with the relative strength of the reflection
continuum, contrary to all simple reflection models.   It is proposed
that continuum-flux correlated changes in the ionization of the
accretion disk surface can explain this spectral variability.   Finally,
we address the issue of X-ray iron line reverberation in the light of
these complicating factors.

\section{Acknowledgements}

The author appreciates support from Hubble Fellowship grant HF-01113.01-98A.
This grant was awarded by the Space Telescope Institute, which is
operated by the Association of Universities for Research in Astronomy,
Inc., for NASA under contract NAS 5-26555.  He also appreciates support
from NASA under LTSA grant NAG5-6337, and the National Science
Foundation under grants AST-9529170 and AST-9876887.


\begin{references}
\reference Alexander T., 1997, in Astronomical Time Series, ed. D.~Maoz
et al., (Dordrecht: Kluwer), 163
\reference Chiang J. et al., 2000, ApJ, 528, 292
\reference Fabian A.~C., Rees M.~J., Stellar L., White N.~E., 1989, MNRAS,
238, 729
\reference Fabian A.~C. et al. 1995, MNRAS, 277, L11
\reference George I.~M., Fabian A.~C., 1991, MNRAS, 249, 352
\reference Guainazzi M. et al., 1999, A\&A, 341, L27
\reference Ichimaru S., 1977, ApJ, 214, 840
\reference Lasota J.-P., Abramowicz M.~A., Chen X., Krolik J., Narayan R.
   1996, ApJ, 462, 142
\reference Lee J., Fabian A.~C., Reynolds C.~S., Brandt W.~N., Iwasawa
K., 2000, MNRAS, in press
\reference Lee J., Fabian A.~C., Brandt W.~N., Reynolds C.~S., Iwasawa
K., 1999, MNRAS, 310, 973
\reference Misra R., 1999, IUCAA preprint 32/99
\reference Misra R., Kembhavi A.~K., 1998, ApJ, 499, 205
\reference Misra R., Sutaria F.~K., 1999, ApJ, 517, 661
\reference Nandra K., George I.~M., Mushotzky R.~F., Turner T.~J., Yaqoob
T., 1997, ApJ, 477, 602
\reference Nayakshin S., Kazanas D., Kallman T.~R., 2000, ApJ, submitted (astro-ph/9909359)
\reference Narayan R., Yi I., 1995, ApJ, 452, 710
\reference Narayan R., Yi I., 1994, ApJ, 428, L13
\reference Novikov I.~D.,  Thorne K.~S., 1973, in {\it Black Holes}, eds
  C.~DeWitte \& B.~S.~DeWitte (Gordon and Breach Science Publicaters, New
  York), P.344
\reference Press W.~H., Rybicki G.~B., Hewitt J.~N., 1992, ApJ, 385, 404
\reference Rees M.~J., 1982, in Riegler G., Blandford R.~D., eds, The
Galactic Center, Am. Inst. Phys., New York, p.166.
\reference Reynolds C.~S., 1996, PhD thesis, University of Cambridge
\reference Reynolds C.~S., 2000, ApJ, 533, 811
\reference Reynolds C.~S., Wilms J., 2000, ApJ,  533, 821
\reference Reynolds C.~S., Fabian A.~C., Inoue H., 1995,  MNRAS, 276, 1311
\reference Reynolds C.~S., Nowak M.~A., Maloney P.~R., 2000, ApJ, in
press (astro-ph/0004068)
\reference Reynolds C.~S., Young A.~J., Begelman M.~C., Fabian A.~C.,
1999, ApJ, 514, 164
\reference Ross R.~R., Fabian A.~C., 1993, MNRAS, 261, 74
\reference Ruszkowski M., 2000, MNRAS, 315, 1
\reference Ruszkowski M., Fabian A.~C., 2000, MNRAS, submitted
\reference Shakura N.~L., Sunyaev R.~A., 1973, A\&A, 24, 337
\reference Skibo, J.G., 1997, ApJ, 478, 522
\reference Tanaka Y. et al.,  1995, Nat, 375, 659
\reference Young A.~J., Reynolds C.~S., 2000, ApJ, 529, 101
\end{references}
\end{document}